\documentclass[aps,prl,twocolumn,superscriptaddress]{revtex4-1}
\usepackage{graphicx}
\usepackage{dcolumn}
\usepackage{bm}
\usepackage{color}
\usepackage{ulem}
\usepackage[usenames,dvipsnames,svgnames,table]{xcolor}

\newcommand{\pstar}{$p^{\star}$}

\newcommand{\pcdw}{$p_{\rm CDW}$}

\newcommand{\RH}{$R_{\rm H}$}
\newcommand{\nH}{$n_{\rm H}$}

\newcommand{\Tc}{$T_{\rm c}$}
\newcommand{\Tstar}{$T^{\star}$}

\newcommand{\Tcdw}{$T_{\rm CDW}$}
\newcommand{\Tmax}{$T_{\rm max}$}

\newcommand{\Hc}{$H_{\rm c2}$}

\begin{document}

%\preprint{}


\title{Thermopower across the phase diagram of the cuprate La$_{1.6-x}$Nd$_{0.4}$Sr$_x$CuO$_4$ :\\
signatures of the pseudogap and charge-density-wave phases}

\author{C.~Collignon$^{\star \star}$}
\email[]{Present address: Department of Physics, Massachusetts Institute of Technology, Cambridge, Massachusetts 02139, USA}
\affiliation{Institut Quantique, D\'epartement de physique \& RQMP, Universit\'e de Sherbrooke, Sherbrooke, Qu\'ebec J1K 2R1, Canada}

\author{A.~Ataei$^{\star \star}$}
\affiliation{Institut Quantique, D\'epartement de physique \& RQMP, Universit\'e de Sherbrooke, Sherbrooke, Qu\'ebec J1K 2R1, Canada}

\author{A.~Gourgout}
\affiliation{Institut Quantique, D\'epartement de physique \& RQMP, Universit\'e de Sherbrooke, Sherbrooke, Qu\'ebec J1K 2R1, Canada}

\author{S.~Badoux}
\affiliation{Institut Quantique, D\'epartement de physique \& RQMP, Universit\'e de Sherbrooke, Sherbrooke, Qu\'ebec J1K 2R1, Canada}

\author{M.~Lizaire}
\affiliation{Institut Quantique, D\'epartement de physique \& RQMP, Universit\'e de Sherbrooke, Sherbrooke, Qu\'ebec J1K 2R1, Canada}

\author{A.~Legros}
\affiliation{Institut Quantique, D\'epartement de physique \& RQMP, Universit\'e de Sherbrooke, Sherbrooke, Qu\'ebec J1K 2R1, Canada}
\affiliation{SPEC, CEA, CNRS-UMR 3680, Université Paris-Saclay, Gif sur Yvette Cedex, France}

\author{S.~Licciardello}
\affiliation{High Field Magnet Laboratory (HFML-EMFL) \& Institute for Molecules and Materials, Radboud University, Toernooiveld 7, 6525 ED Nijmegen, The Netherlands}

\author{S.~Wiedmann}
\affiliation{High Field Magnet Laboratory (HFML-EMFL) \& Institute for Molecules and Materials, Radboud University, Toernooiveld 7, 6525 ED Nijmegen, The Netherlands}

\author{J.-Q.~Yan}
\affiliation{Materials Science and Engineering Program, Mechanical Engineering, University of Texas at Austin, Austin, Texas 78712, USA}

\author{J.-S.~Zhou}
\affiliation{Materials Science and Engineering Program, Mechanical Engineering, University of Texas at Austin, Austin, Texas 78712, USA}

\author{Q.~Ma}
\affiliation{Department of Physics and Astronomy, McMaster University, Hamilton, Ontario L8S 4M1, Canada}

\author{B.~D.~Gaulin}
\affiliation{Department of Physics and Astronomy, McMaster University, Hamilton, Ontario L8S 4M1, Canada}
\affiliation{Canadian Institute for Advanced Research, Toronto, Ontario M5G 1M1, Canada}

\author{Nicolas~Doiron-Leyraud}
\email[]{nicolas.doiron-leyraud@usherbrooke.ca}
\affiliation{Institut Quantique, D\'epartement de physique \& RQMP, Universit\'e de Sherbrooke, Sherbrooke, Qu\'ebec J1K 2R1, Canada}

\author{Louis~Taillefer}
\email[]{louis.taillefer@usherbrooke.ca}
\affiliation{Institut Quantique, D\'epartement de physique \& RQMP, Universit\'e de Sherbrooke, Sherbrooke, Qu\'ebec J1K 2R1, Canada}
\affiliation{Canadian Institute for Advanced Research, Toronto, Ontario M5G 1M1, Canada}

\date{\today}

\begin{abstract}

The Seebeck coefficient (thermopower) $S$ of the cuprate superconductor La$_{1.6-x}$Nd$_{0.4}$Sr$_x$CuO$_4$ was measured across its doping phase diagram (from $p = 0.12$ to $p = 0.25$), at various temperatures down to $T \simeq 2$~K, in the normal state accessed by suppressing superconductivity with a magnetic field up to $H = 37.5$~T.
The magnitude of $S/T$ in the $T=0$ limit is found to suddenly increase, by a factor $\simeq 5$, when the doping is reduced below \pstar~$=0.23$, the critical doping for the onset of the pseudogap phase.
This confirms that the pseudogap phase causes a large reduction of the carrier density $n$, consistent with a drop from $n = 1 + p$ above \pstar~to $n = p$ below \pstar, as previously inferred from measurements of the Hall coefficient, resistivity and thermal conductivity.
When the doping is reduced below $p = 0.19$, a qualitative change is observed whereby $S/T$ decreases as $T \to 0$, eventually to reach negative values at $T=0$.
In prior work on other cuprates, negative values of $S/T$ at $T \to 0$ were shown to result from a reconstruction of the Fermi surface caused by charge-density-wave (CDW) order.
We therefore identify \pcdw~$\simeq 0.19$ as the critical doping beyond which there is no CDW-induced Fermi surface reconstruction.
The fact that \pcdw~is well separated from \pstar~reveals that there is a doping range below \pstar~where the transport signatures of the pseudogap phase are unaffected by CDW correlations, as previously found in YBa$_2$Cu$_3$O$_y$ and La$_{2-x}$Sr$_x$CuO$_4$. 

\end{abstract}


\maketitle

%%%%%%%%%%%%%%%%%%%%%%%%%%%%%%%%%%%%%%%%%%%%%%%%%%%%%%%%%%%%
%%%%%%%%%%%%%%%%%%%%    INTRODUCTION  %%%%%%%%%%%%%%%%%%%%%%%%%%%%%
%%%%%%%%%%%%%%%%%%%%%%%%%%%%%%%%%%%%%%%%%%%%%%%%%%%%%%%%%%%%

\section{Introduction}
The chief mystery in our understanding of cuprate superconductors is the pseudogap phase~\cite{keimer2015},
a region in the temperature-doping phase diagram of hole-doped cuprates bounded by the crossover temperature \Tstar~
and the critical doping \pstar, where the pseudogap phase ends at $T=0$~\cite{cyr-choiniere2018}.
Some of the sharpest experimental signatures of this phase have recently been found in the $T=0$ limit,
as the doping $p$ is tuned across \pstar,
in the absence of superconductivity, removed by applying a large magnetic field~\cite{proust2019}.
A key signature is the drop in the Hall number \nH, from \nH~$\simeq 1 + p$ above \pstar~to \nH~$\simeq p$ below \pstar,
as revealed by measurements of the Hall effect in YBa$_2$Cu$_3$O$_y$ (YBCO)~\cite{badoux2016}, 
La$_{1.6-x}$Nd$_{0.4}$Sr$_x$CuO$_4$ (Nd-LSCO)~\cite{collignon2017} and Bi$_2$Sr$_2$CuO$_6$ (Bi2201)~\cite{lizaire2020}.
It is tempting to equate the Hall number \nH~with the actual carrier density $n$, as in the case of simple single-band metals.
If this were correct for cuprates, then a key property of the pseudogap phase is elucidated:
it causes the carrier density to change from $n = 1 + p$ to $n = p$. 
The natural explanation for such a change is the onset of a state that breaks translational symmetry
by introducing a new periodicity -- e.g. with wavevector ${\bf Q} = (\pi,\pi)$ --
which causes a transformation of the Fermi surface from a large cylinder (of volume proportional to $1+p$)
to small nodal hole pockets (of volume proportional to $p$)~\cite{storey2016}.

However, before we can assert that these are amongst the defining properties of the pseudogap phase, further evidence for a drop in carrier density is needed, because other mechanisms can also cause \nH~to drop.
For example, a change in the curvature of the Fermi surface will affect \nH,
as shown in the case of a nematic transition, for which \nH~can mimic the behavior found in cuprates across \pstar~\cite{maharaj2017}.
The loss of quasiparticle coherence (or the advent of additional scattering) has also been invoked to explain the drop in \nH, as opposed to a transformation from large Fermi surface to small~\cite{putzke2019}.

In Nd-LSCO, the observation of a large increase in the electrical resistivity~\cite{collignon2017,daou2009} upon entering the pseudogap phase, along with a correspondingly large decrease in the electronic thermal conductivity~\cite{michon2018},
rules out a simple change in the curvature of the Fermi surface as the explanation for the change in \nH.
Indeed, for the conductivity to drop, either the carrier density has to drop or the scattering rate has to increase (or both)\textcolor{black}{~\cite{storey2017}}.

%%%%%%%%%%%%%%% Begin Figure 1 %%%%%%%%%%%%%%%%%%%%%%%%%%%%%%%%%%%%%%%%%%%%%%%%%%%%%%%%%%%%
%
\begin{figure}[t!]
\includegraphics[width=0.48\textwidth]{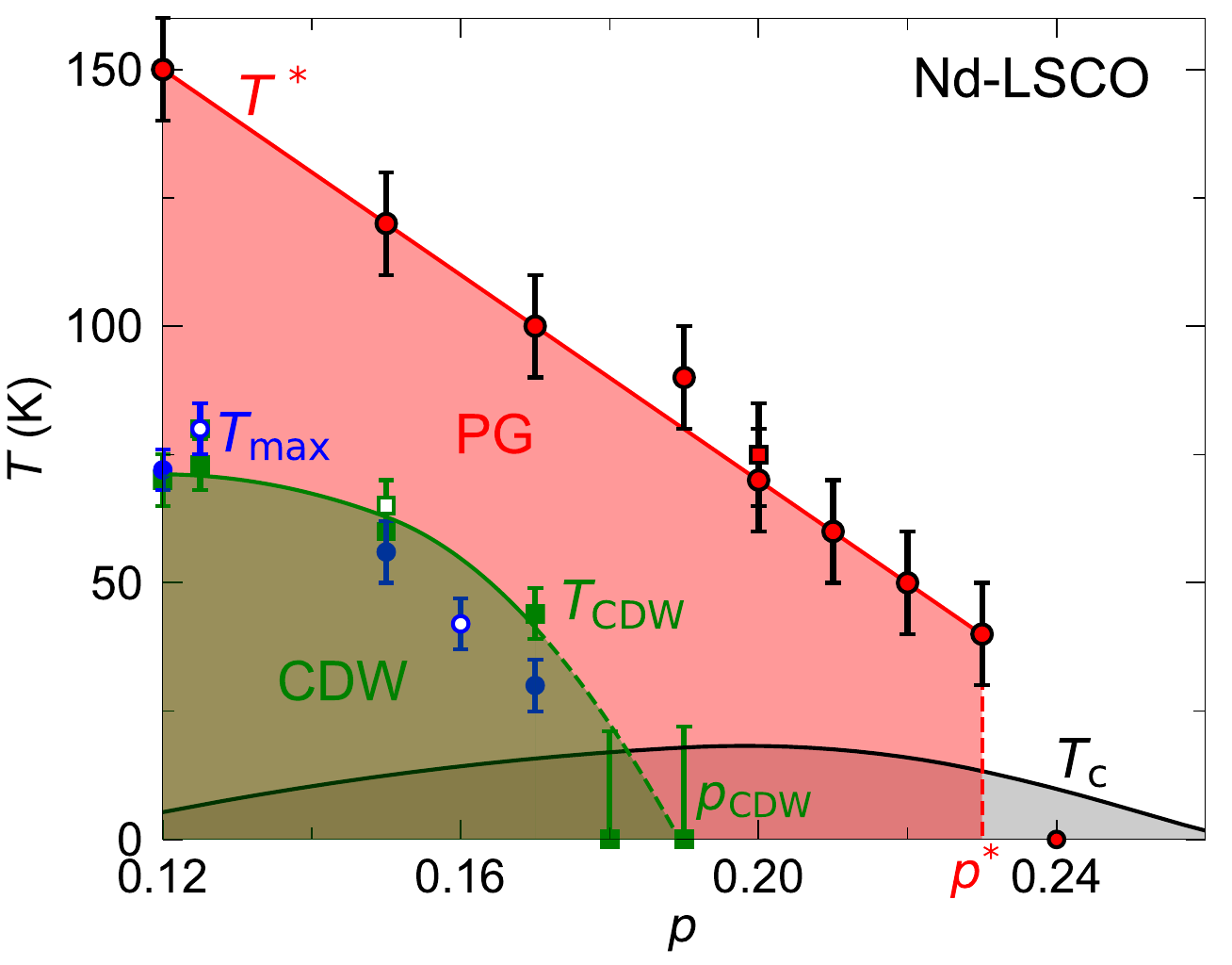}
\caption{
Temperature-doping phase diagram of Nd-LSCO showing 
the pseudogap temperature \Tstar~extracted from resistivity (red dots, from~\cite{collignon2017} 
and Fig.~S1 in the Supplementary Material) 
and ARPES measurements (red square, from~\cite{matt2015}), 
the CDW ordering temperature \Tcdw~as seen in  
x-ray diffraction measurements (green squares, from\textcolor{black}{~\cite{zimmermann1998,niemoller1999,gupta2020}}), 
the temperature \Tmax~of the maximum in $S/T$ vs $T$ (blue dots, see main text and Fig.~\ref{Fig6}) \textcolor{black}{interpreted as signaling the onset of CDW order}, 
and a schematic of the zero-field superconducting transition temperature \Tc~(black line). 
Corresponding values for \Tmax~\cite{laliberte2011} and \Tcdw~\cite{fink2011} in 
La$_{1.8-x}$Eu$_{0.2}$Sr$_x$CuO$_4$ (Eu-LSCO) are shown as open symbols.
The red and \textcolor{black}{green} full lines are guides to the eye.
The red dashed line marks the end of the pseudogap phase, at the critical doping \pstar~$= 0.23 \pm 0.01$~\cite{collignon2017}.
The \textcolor{black}{green} dashed line is a linear extension of the full \textcolor{black}{green} line, extrapolating to $p =$~\pcdw~$\simeq 0.19$ at $T=0$.
}
\label{Fig1}
\end{figure}
%
%%%%%%%%%%%%%%% End Figure 1 %%%%%%%%%%%%%%%%%%%%%%%%%%%%%%%%%%%%%%%%%%%%%%%%%%%%%%%%%%%%

Here, we use a new experimental probe to investigate the pseudogap critical point, namely the thermopower, or Seebeck coefficient $S$.
Unlike the Hall coefficient, $S$ does not depend on the curvature or shape of the Fermi surface.
Unlike the conductivity (electrical or thermal), it does not depend, at least to first order, on the scattering rate.
Fundamentally, $S$ is the specific heat per carrier.
In the simple case of a free electron gas, the Seebeck coefficient is given by~\cite{behnia2004,miyake2005}:
\begin{equation}\label{eq1}
\frac{S}{T} = \pm \frac{\pi ^2}{3}    \frac{k_B^2}{e}    \frac{1}{n}  N (\epsilon _F) 
\end{equation}
where $e$ is the electron charge and $k_B$ Boltzmann's constant, $N ( \epsilon _F )$ is the density of states at the Fermi energy, and $n$ is the carrier density (the negative sign is for electrons, the positive sign is for holes).
The analogous expression for the electronic specific heat is:
\begin{equation}\label{eq2}
\frac{C_{el}}{T} = \frac{\pi ^2 k_B ^2}{3} N(\epsilon _F).
\end{equation}
Combining the two expressions yields:
\begin{equation}
\frac{S}{T} = \pm \frac{1}{n e} \frac{C_{el}}{T} .
\end{equation}
While seemingly over-simplistic, this relation was shown to hold in the $T=0$ limit for a large variety of materials, even in the presence of strong electronic correlations, including heavy-fermion metals, organic conductors and cuprates~\cite{behnia2004}.
In the cuprate YBCO, measurements in the field-induced normal state yield a large Seebeck coefficient, namely $S/T = - 0.8~\mu$V/K$^2$~\cite{laliberte2011}, and a small specific heat, namely $C_{el}/T = \gamma = 2.4$~mJ/K$^2$mol-Cu~\cite{kacmarcik2018}, in the $T=0$ limit, for $p=0.11$.
From Eq.~3, this implies a very small carrier density, namely $n = 0.032$ per (planar) Cu atom.
In YBCO at $p=0.11$, the Fermi surface is well known from quantum oscillations and the volume of its tiny electron pocket is precisely given by the oscillation frequency, $F = 530$~T, which yields $n = 0.038$~\cite{doiron-leyraud2007}.
This excellent agreement confirms that the Seebeck coefficient can be used as a measure of the carrier density in cuprates.

Here we focus on Nd-LSCO, a single-layer, tetragonal cuprate superconductor with a low critical temperature \Tc~and critical field \Hc, such that superconductivity can be readily suppressed with static fields down to $T \to 0$. 
In Fig.~\ref{Fig1}, we display the temperature-doping phase diagram of Nd-LSCO, showing the pseudogap temperature \Tstar, defined as the temperature at which the resistivity $\rho(T)$ departs from its high-temperature $T$-linear behaviour~\cite{collignon2017,daou2009} (Fig.~\ref{Fig7}(a)), as a function of doping.
At $p = 0.24$, $\rho(T)$ remains $T$-linear down to $T \to 0$ (Fig.~\ref{Fig7}(a);~\cite{collignon2017,daou2009,michon2018}), 
implying that the pseudogap phase ends at \pstar~$= 0.23 \pm 0.01$.
This agrees well with angle-resolved photoemission spectroscopy (ARPES) measurements on Nd-LSCO~\cite{matt2015}, 
which find the opening of an anti-nodal pseudogap below \Tstar~$= 75$~K at $p = 0.20$ (red square, Fig.~\ref{Fig1})
and no pseudogap at $p=0.24$ (down to $T \simeq$~\Tc).
Hall effect measurements in Nd-LSCO find that the Hall number \nH~$\simeq 1 + p$ at $p = 0.24 >$~\pstar~and \nH~$\simeq p$ at $p = 0.20 <$~\pstar~\cite{collignon2017}, in good agreement with YBCO~\cite{badoux2016} and, as reported recently, with Bi2201~\cite{lizaire2020},
pointing to a universal signature.

At $p=0.12$, charge-density-wave (CDW) order appears below a temperature \Tcdw~$=70 \pm 5$~K (Fig.~\ref{Fig1}), as measured by neutron~\cite{tranquada1995} and x-ray~\cite{zimmermann1998} diffraction.
At $p = 0.15$, \Tcdw~$=62 \pm 5$~K~\cite{niemoller1999}.
How the boundary of the CDW phase in Nd-LSCO evolves beyond $p = 0.15$, and in particular at what doping \pcdw~ends, is not known.

In the present Letter, we report measurements of the Seebeck coefficient of Nd-LSCO in the field-induced normal state down to low temperature ($T \simeq 2$~K), accessed by applying magnetic fields up to 37.5~T,
for dopings ranging from $p = 0.12$ to $p = 0.25$, thereby spanning both \pstar~and the likely termination of the CDW phase at $p_{\rm CDW}$.

We find that $S/T$ at $T \to 0$ exhibits a sharp five-fold increase in magnitude upon crossing below \pstar, which confirms -- via Eq.~3 --
that the carrier density drops upon entering the pseudogap phase (given that $C_{el} / T$ decreases below \pstar~\cite{michon2019}).
With decreasing $p$, $S/T$ remains high until $p = 0.19$, at which point it starts to decrease towards negative values.
We therefore find that the usual effect of CDW order on the transport properties of Nd-LSCO are observed up to at most \pcdw~$= 0.19$, which is well below \pstar.

%%%%%%%%%%%%%%% Begin Figure 2 %%%%%%%%%%%%%%%%%%%%%%%%%%%%%%%%%%%%%%%%%%%%%%%%%%%%%%%%%%%%
%
\begin{figure*}[t!]
\centering
\includegraphics[width=0.80\textwidth]{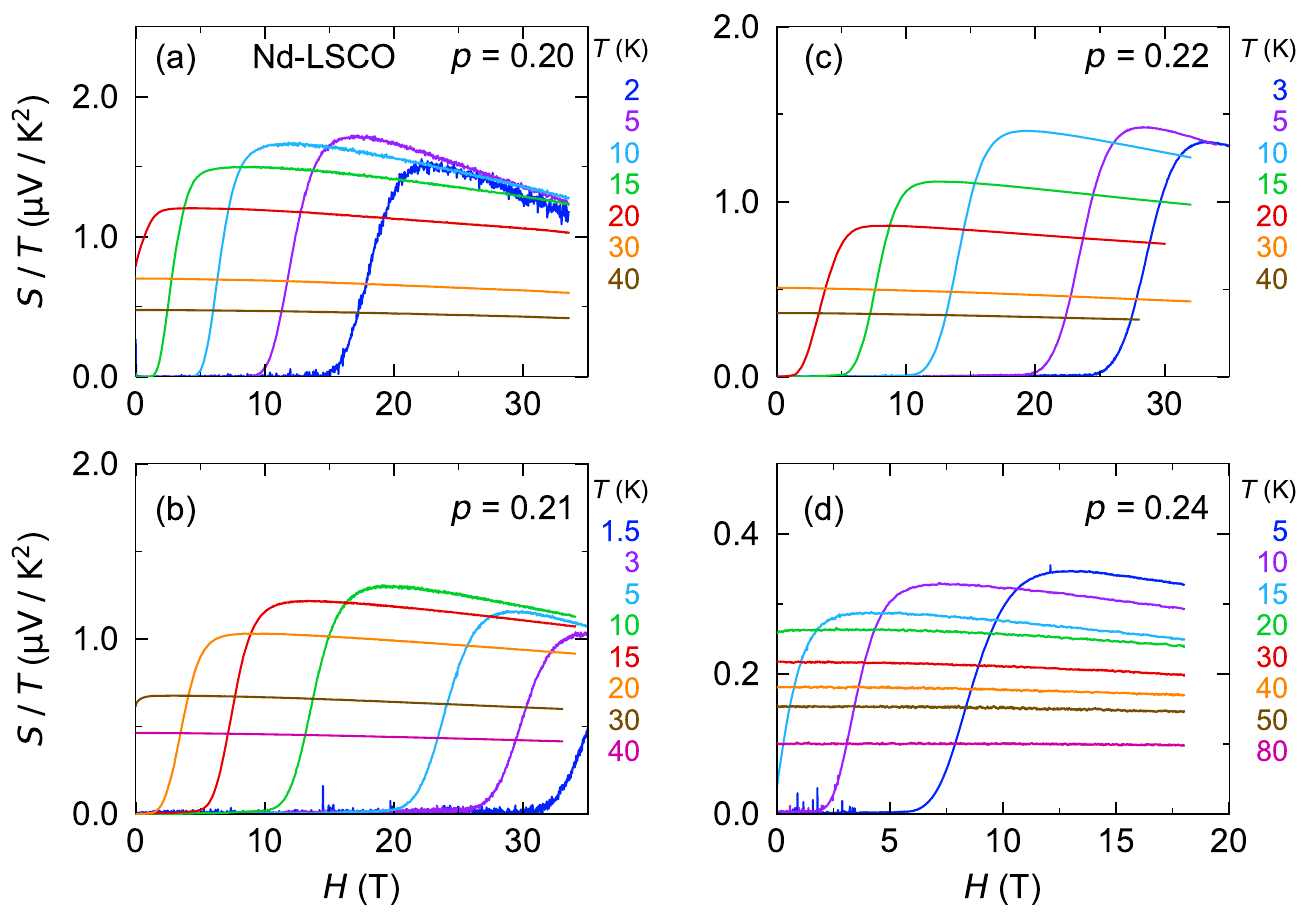}
\caption{
Isotherms of the Seebeck coefficient 
in Nd-LSCO, plotted as $S/T$ vs magnetic field $H$,
at temperatures and dopings as indicated. 
Isotherms for $p = 0.23$ and 0.25 are shown in Fig.~S2 of the Supplementary Material.
Note the change of scale for $S/T$ -- by a factor of about 4 -- when going from $p = 0.20$, 0.21, 0.22 $<$~\pstar~to $p = 0.24 >$~\pstar.
}
\label{Fig2}
\end{figure*}
%
%%%%%%%%%%%%%%% End Figure 2 %%%%%%%%%%%%%%%%%%%%%%%%%%%%%%%%%%%%%%%%%%%%%%%%%%%%%%%%%%%%

\section{Methods}
The thermopower was measured in ten single-crystalline samples of Nd-LSCO, each with a different doping $p$: 0.12, 0.15, 0.17, 0.19, 0.20, 0.21, 0.22, 0.23, 0.24 and 0.25.
Five of those samples -- with $p = 0.20$, 0.21, 0.22, 0.23 and 0.24 -- were previously studied with measurements of
resistivity and Hall effect~\cite{collignon2017}.
Details of the sample and contact preparation can be found there.
The thermal conductivity of those same 5 samples was reported in~\cite{michon2018}, along with another two -- with $p = 0.12$ and 0.15.
Structural details on some of our Nd-LSCO samples, in particular where they lie on the structural phase diagram, can be found in ref.~\cite{dragomir2020}.
The value of \Tc~(defined as the point of zero resistance in zero field) varies from a local minimum of \Tc~$=5$~K at $p=0.12$ to 
a local maximum of \Tc~$\simeq 15$~K at $p \simeq 0.18$ back down to \Tc~$=5$~K at $p=0.25$ (see Extended Data Fig.~2b in~\cite{michon2019}).
Samples were cut from large single crystals of Nd-LSCO grown by a traveling float-zone technique, 
into small rectangular platelets of typical dimensions 1 mm x 0.5 mm x 0.2 mm, with the shortest dimension along the $c$~axis.
%

%%%%%%%%%%%%%%% Begin Figure 3 %%%%%%%%%%%%%%%%%%%%%%%%%%%%%%%%%%%%%%%%%%%%%%%%%%%%%%%%%%%%
%
\begin{figure}[t!]
\includegraphics[width=0.47\textwidth]{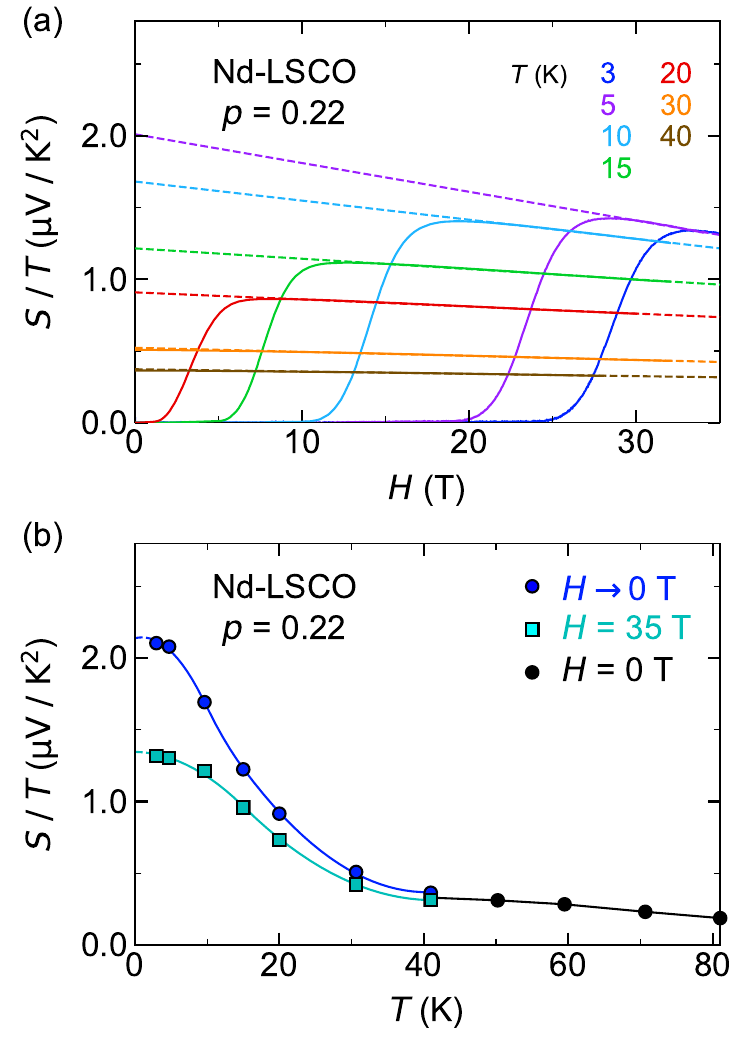}
\caption{
(a) 
Isotherms of $S/T$ versus $H$ for Nd-LSCO at $p = 0.22$, at temperatures as indicated. 
The dashed lines are linear-$H$ fits to $S/T$ in the normal state above the superconducting transition, highlighting the negative field dependence.
In order to remove this magneto-Seebeck effect, we back extrapolate the fits to $H = 0$ and thereby extract the zero-field values of $S/T$ in the absence of superconductivity. 
Data and fits for all other dopings are shown in Fig.~S2 of the Supplementary Material. 
(b) $S/T$ versus $T$ for Nd-LSCO at $p = 0.22$, showing data taken in zero field above $T_c$ (black dots), 
at $H$ = 35~T (turquoise dots), 
and the values extrapolated to $H \rightarrow 0$ (blue dots, obtained from fits in panel (a)). 
All lines are a guide to the eye.
(Data at $H = 33$~T for all dopings are shown in Fig.~S3 of the Supplementary Material.)
}
\label{Fig3}
\end{figure}
%
%%%%%%%%%%%%%%% End Figure 3 %%%%%%%%%%%%%%%%%%%%%%%%%%%%%%%%%%%%%%%%%%%%%%%%%%%%%%%%%%%%

We employ a standard one-heater-two-thermometers steady-state method to apply and measure a temperature difference $\Delta T_x$ along the length of our samples.
At their cold end, samples were anchored to a copper block which provides both an electrical and thermal ground. The heat current was applied in the $ab$ plane of the low-temperature tetragonal structure of Nd-LSCO.
The longitudinal (Seebeck) voltage $\Delta V_x$ was measured on the very same contacts as used for measuring $\Delta T_x$, which removes the uncertainty on the geometrical factor.
The Seebeck coefficient $S$ is then simply given by $S = \Delta V_x/\Delta T_x$.
For measurements in a magnetic field, the field was applied along the $c$ axis of the crystal structure of Nd-LSCO. The $ab$-plane Seebeck voltage was measured at positive and negative polarities of the field, and was symmetrized with respect to field in order to obtain the pure Seebeck signal, free of contamination from the Nernst effect.
Measurements in fields up to $H$ = 18~T were performed at Sherbrooke, and up to 37.5~T at the High Field Magnet Laboratory (HFML) in Nijmegen.

The thermopower of Nd-LSCO was previously measured on polycrystalline samples of La$_{2-y-x}$Nd$_{y}$Sr$_x$CuO$_4$ with $y=0.6$, in zero field,
for $p = 0.08$, 0.10, 0.11, 0.12, 0.13, 0.15, 0.17, 0.19 and 0.21, as reported in ref.~\onlinecite{hucker1998},
and on single-crystal samples of La$_{2-y-x}$Nd$_{y}$Sr$_x$CuO$_4$ with $y=0.4$, in fields up to 15~T, for $p = 0.20$ and 0.24, as reported in ref.~\onlinecite{daou_thermo2009}.
The data reported here are in excellent agreement with these prior low-field data.

\section{Results}
In Fig.~\ref{Fig2},
we show isothermal field sweeps of the Seebeck coefficient $S$, 
expressed as $S/T$ vs $H$,
at $p = 0.20$, 0.21, 0.22, and 0.24. 
For all dopings, $S/T$ is null in the superconducting state at low field, rises quickly upon crossing the vortex-solid transition and then reaches the normal-state value at $H_{c2}$ and above. 
At all dopings we observe an increase of the normal-state $S/T$ with decreasing temperature. 
For a plot of $S/T$ vs $T$ at $H = 35$~T, for $p = 0.22$, see Fig.~\ref{Fig3}(b).
(For a similar plot at other dopings, see Fig.~S2 of the Supplementary Material.)
In Fig.~\ref{Fig2},
we readily observe from the raw data at 5~K and at the highest field that the amplitude of $S/T$ remains roughly constant as a function of doping for $p = 0.20, 0.21, 0.22 <$~\pstar, 
but then drops by a factor of about 4 when going to $p = 0.23, 0.24, 0.25 >$~\pstar.
This is one of the main findings of our study: the pseudogap phase causes a large increase in $S/T$ in the $T = 0$ limit, in the ground state without superconductivity.
This increase is strong evidence for a drop in carrier density, given the fundamental relation 
$S/T \propto (C_{\rm el}/T) / n$ (Eq.~3), since $C_{\rm el}/T$ decreases below \pstar~\cite{michon2019}.

%%%%%%%%%%%%%%% Begin Figure 4 %%%%%%%%%%%%%%%%%%%%%%%%%%%%%%%%%%%%%%%%%%%%%%%%%%%%%%%%%%%%
%
\begin{figure}[t!]
\includegraphics[width=0.45\textwidth]{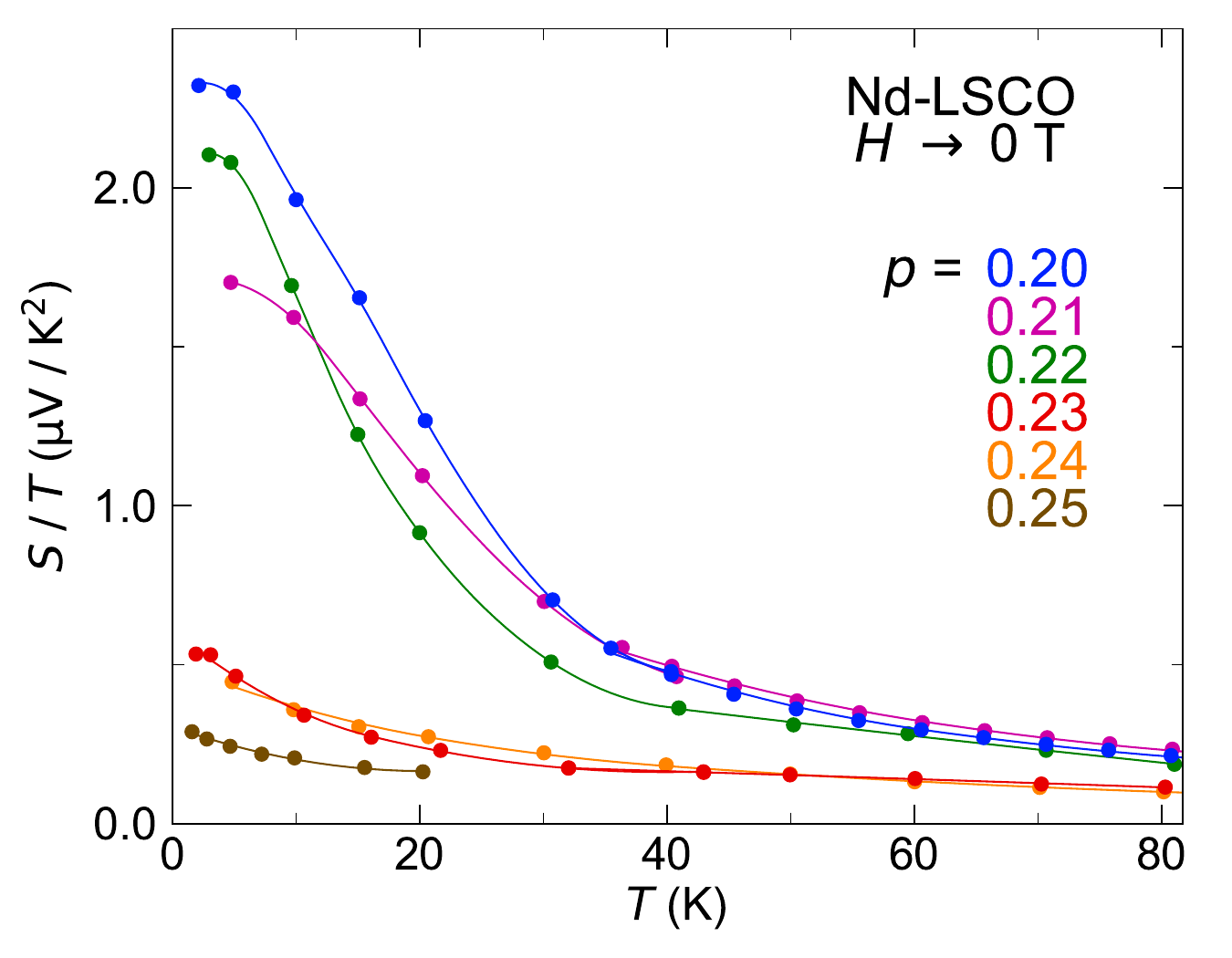}
\caption{
$S/T$ versus $T$ extrapolated to $H \rightarrow 0$, at dopings as indicated. 
The data points are based on the extrapolations shown in Fig.~\ref{Fig3}(a) and in Fig.~S2 of the Supplementary Material. 
There is a clear and abrupt change in the magnitude of $S/T$ at $T \simeq 0$ upon crossing into the pseudogap phase ($p < 0.23$).
All lines are a guide to the eye.
}
\label{Fig4}
\end{figure}
%
%%%%%%%%%%%%%%% End Figure 4 %%%%%%%%%%%%%%%%%%%%%%%%%%%%%%%%%%%%%%%%%%%%%%%%%%%%%%%%%%%%

%
In Fig.~\ref{Fig2}, we see that
the normal-state $S/T$ displays a negative field dependence.
This is a bit like the magneto-resistance seen in the same samples~\cite{collignon2017},
which causes the conductivity to decrease with increasing field.
This magneto-Seebeck effect is more pronounced in our sample with $p = 0.20$ than in our sample with $p = 0.24$,
and it is nearly absent in our samples with $p = 0.23$ and $p = 0.25$ (see Fig.~S2 of the Supplementary Material.).
In order to correct for this effect, and obtain the value of $S/T$ free from this field-induced decrease,
we back-extrapolate the isotherms to $H = 0$ as illustrated in Fig.~\ref{Fig3}(a).
As shown by the straight dashed lines for $p = 0.22$ (and in Fig.~S2 of the Supplementary Material for other dopings), the field dependence of $S/T$ in the normal state is well described by a linear-$H$ dependence. 
We apply linear fits to our data, extend them below \Hc, and take the $y$-intercept of those fits at $H = 0$ to obtain the value of $S/T (H \rightarrow 0)$.
In Fig.~\ref{Fig3}(b), we plot the resulting curve of $S/T (H \rightarrow 0)$ vs $T$ and also our raw curve of $S/T (H = 35~T)$ vs $T$. 
Both continuously extend the zero-field curve taken above $T_c$, but they clearly separate at low temperature, where $S/T (H \rightarrow 0)$ saturates at a significantly higher value, 
revealing the magnitude of the magneto-Seebeck effect in that sample.

Applying the same analysis to $p = 0.20$, 0.21, 0.23, 0.24, and 0.25 (see in Fig.~S2 of the Supplementary Material), we plot the resulting curves of $S/T (H \rightarrow 0)$ vs $T$ in Fig.~\ref{Fig4}. 
This immediately reveals a clear thermopower signature of the pseudogap phase: upon crossing below \pstar~$= 0.23$, $S/T (H \rightarrow 0)$ at low temperature undergoes a large and sudden increase. 
The curves of $S/T (H \rightarrow 0)$ essentially appear to belong to one of two groups depending on whether they are above or below \pstar, with little variation within each group, clear evidence that the sudden jump is caused by the onset of the pseudogap phase at \pstar. 
At low temperature, the value of $S/T (H \rightarrow 0)$ increases by a factor of about 5 in going from $p >$~\pstar~to $p <$~\pstar.
The same factor is observed in the raw data at high fields (see Fig.~\ref{Fig6}(b) and Fig.~S3 of the Supplementary Material).

%%%%%%%%%%%%%%% Begin Figure 5 %%%%%%%%%%%%%%%%%%%%%%%%%%%%%%%%%%%%%%%%%%%%%%%%%%%%%%%%%%%%
%
\begin{figure}[t!]
\includegraphics[width=0.44\textwidth]{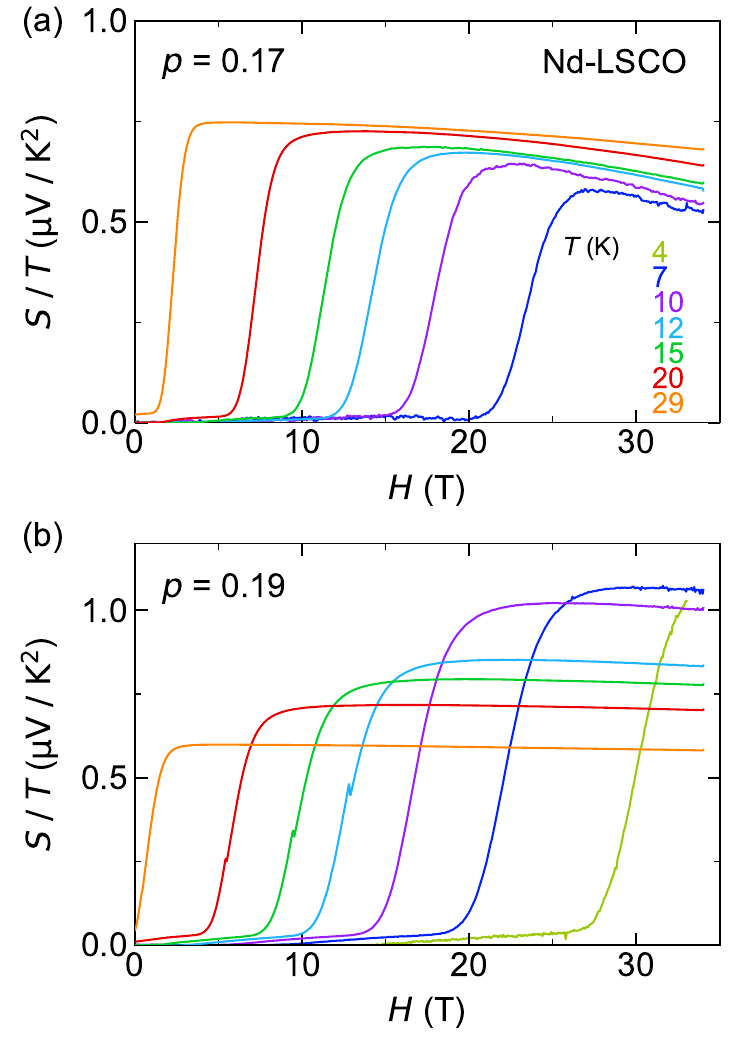}
\caption{
Isotherms of the Seebeck coefficient plotted as $S/T$ vs $H$ for Nd-LSCO at $p = 0.17$ (a) and $p = 0.19$ (b), at temperatures as indicated. 
Note that there is a reversal of the temperature dependence between the two dopings, with $S/T$ now decreasing with decreasing temperature for $p = 0.17$.
\textcolor{black}{For $p = 0.19$, the 4 K isotherm will likely saturate above the 7 K isotherm once the normal state is reached (at higher fields), showing that \Tmax~must be below 4~K.}
}
\label{Fig5}
\end{figure}
%
%%%%%%%%%%%%%%% End Figure 5 %%%%%%%%%%%%%%%%%%%%%%%%%%%%%%%%%%%%%%%%%%%%%%%%%%%%%%%%%%%%
%
%%%%%%%%%%%%%%% Begin Figure 6 %%%%%%%%%%%%%%%%%%%%%%%%%%%%%%%%%%%%%%%%%%%%%%%%%%%%%%%%%%%%
%
\begin{figure}[t!]
\includegraphics[width=0.46\textwidth]{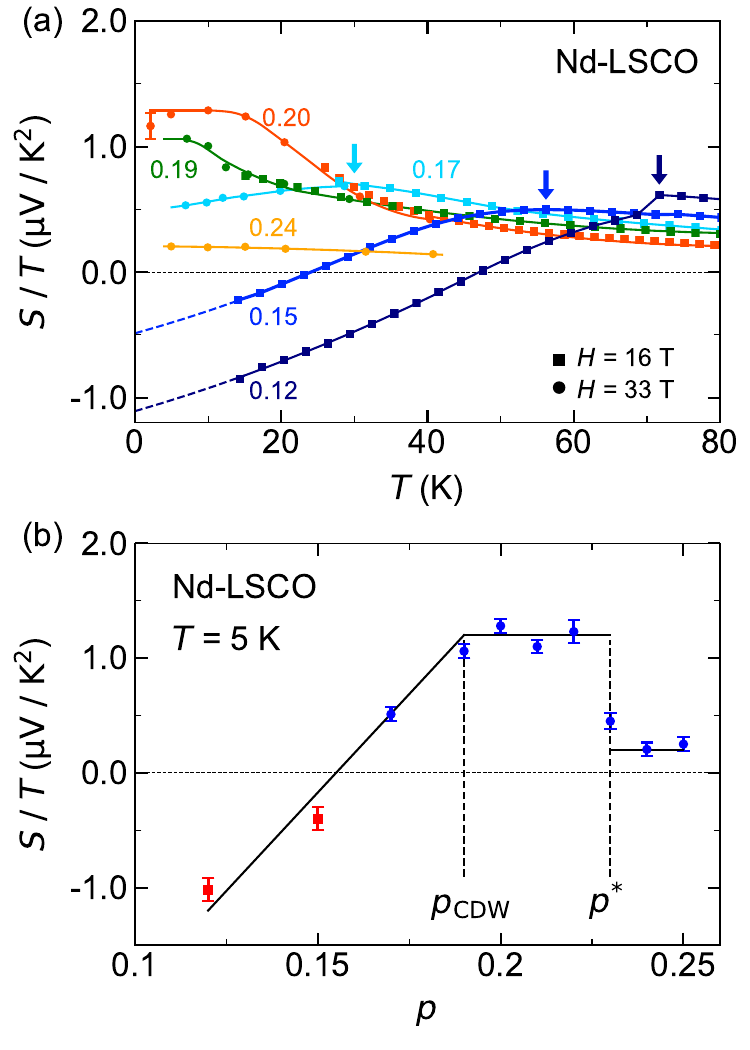}
\caption{
(a) 
$S/T$ as a function of temperature for Nd-LSCO at six dopings as indicated, in a field of $H = 16$~T (squares) and 33~T (dots). 
We observe a clear evolution of $S/T$ vs $T$ with doping: for $p = 0.24 >$~\pstar, $S/T$ has a weak, nearly featureless temperature dependence; then for $p = 0.19$ and $0.20 <$~\pstar,
$S/T$ exhibits a pronounced growth at low $T$, in the pseudogap phase below \Tstar;
and finally for $p < 0.19$, $S/T$ shows a clear drop at low temperature, which results in a negative $S/T$ at $p = 0.12$ and 0.15. 
The arrows indicate the temperature of the maximum in $S/T$ vs $T$, labeled \Tmax~and plotted as a function of doping in Fig.~\ref{Fig1}. 
(b) 
$S/T$ taken in the normal state at $T = 5$~K, in a field $H = 16$~T 
(red squares) and $H = 33$~T (blue dots), 
as a function of doping. 
This plot of $S/T$ vs $p$ in the $T = 0$ limit shows clearly the sudden (5-fold) jump in $S/T$ upon entry into the pseudogap phase at \pstar~$= 0.23$, 
followed by the gradual fall to negative values, a transport signature of entry into the CDW phase at \pcdw~$\simeq 0.19$, or slightly below.
}
\label{Fig6}
\end{figure}
%
%%%%%%%%%%%%%%% End Figure 6 %%%%%%%%%%%%%%%%%%%%%%%%%%%%%%%%%%%%%%%%%%%%%%%%%%%%%%%%%%%%

We now turn to dopings $p < 0.20$ and in Fig.~\ref{Fig5} we show isotherms of $S/T$ for Nd-LSCO at $p = 0.19$ and 0.17. 
For $p = 0.19$, the data are consistent with those at $p = 0.20$, having a similar amplitude and showing an increase with decreasing temperature. 
At $p = 0.17$, however, one important change occurs: at low temperatures, $S/T$ decreases with decreasing temperature. 
The full temperature dependence of $S/T$ is displayed in Fig.~\ref{Fig6}(a), 
where we see that $S/T$ vs $T$ at $p = 0.17$ goes through a maximum at $T \simeq 28$~K, before falling at low temperature. 
This is a clear departure from the low-temperature enhancement seen for $0.19 \leq p < 0.23$. 
As shown in Fig.~\ref{Fig6}(a), reducing the doping further to $p = 0.15$ and 0.12 shifts the maximum in $S/T$ to higher temperatures and makes the downturn more pronounced, resulting in negative values of $S/T$ as $T \rightarrow 0$. 
When plotted as a function of doping (Fig.~\ref{Fig6}(b)), the low-temperature value of $S/T$ is seen to start its drop towards negative values at $p \simeq 0.19$, or slightly below.

A negative Seebeck coefficient was previously reported for a number of cuprates near $p \simeq 1/8$, namely  
La$_{2-x}$Sr$_x$CuO$_4$ (LSCO)~\cite{badoux2016a},
La$_{2-x}$Ba$_x$CuO$_4$ (LBCO)~\cite{li2007}, 
Nd-LSCO~\cite{nakamura1992,hucker1998},
Eu-LSCO~\cite{hucker1998,chang2010}, 
YBCO~\cite{chang2010,laliberte2011}, 
and
HgBa$_2$CuO$_{4 + \delta}$(Hg1201)~\cite{doiron-leyraud2013}.
These studies showed that a negative Seebeck coefficient is a consequence of the Fermi surface reconstruction caused by the CDW order. 
Our extensive data on Nd-LSCO now reveal over what doping range this reconstruction is present in that material.

%%%%%%%%%%%%%%% Begin Figure 7 %%%%%%%%%%%%%%%%%%%%%%%%%%%%%%%%%%%%%%%%%%%%%%%%%%%%%%%%%%%%
%
\begin{figure}[t!]
\includegraphics[width=0.42\textwidth]{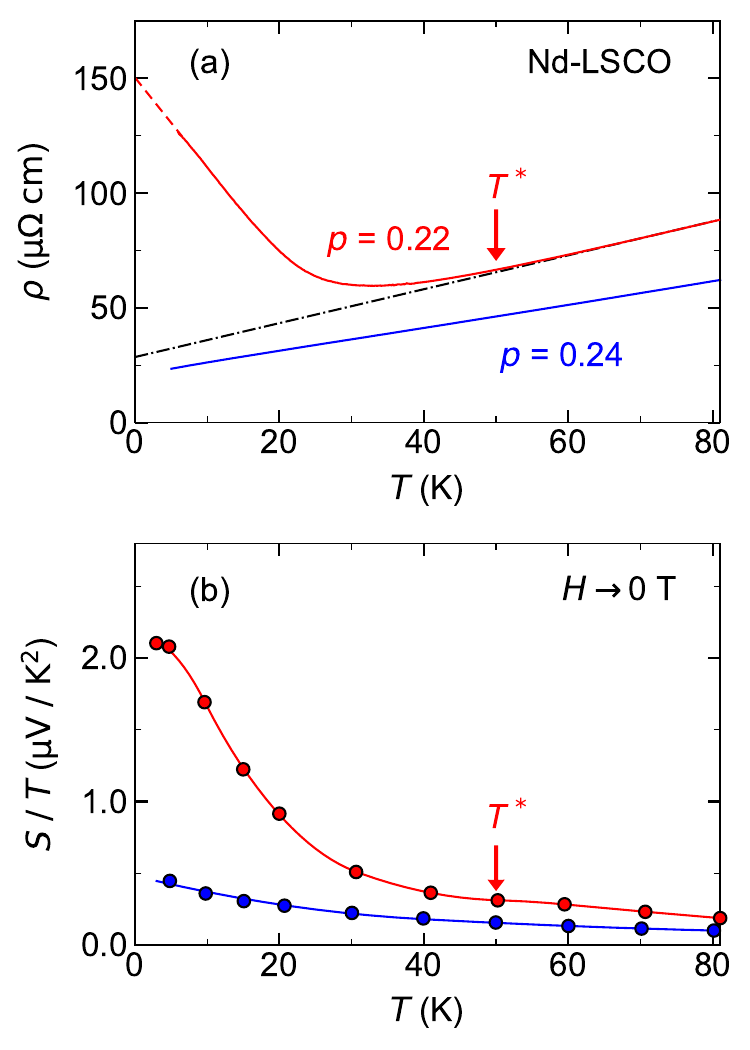}
\caption{
(a) 
Resistivity $\rho(T)$ as a function of $T$ in the field-induced normal state of Nd-LSCO at $H = 33$~T, at dopings as indicated
(from~\cite{collignon2017}). 
At $p = 0.24$ (blue),
there is no pseudogap and the resistivity remains linear down to $T \rightarrow 0$. 
At $p = 0.22$ (red), 
the pseudogap opening causes an upward deviation from linearity (dash-dotted line) below \Tstar, 
and a large upturn in resistivity at low temperature. 
The red dashed line is a linear extrapolation to $T=0$ of the data at $p=0.22$ in 33~T; it yields the zero-temperature estimate of
$\rho(0)$.
The dash-dotted line yields the residual resistivity $\rho_0$ in the absence of the pseudogap (see text),
a measure of the disorder level in the sample.
(b) 
$S/T$ as a function of $T$ based on extrapolated $H \rightarrow 0$ data,
at the same two dopings. 
In analogy to the resistivity, the pseudogap at $p = 0.22$ causes a large enhancement of $S/T$ below $T^{\star}$ with respect to the otherwise slowly varying curve seen in the absence of the pseudogap at $p = 0.24$.
}
\label{Fig7}
\end{figure}
%
%%%%%%%%%%%%%%% End Figure 7 %%%%%%%%%%%%%%%%%%%%%%%%%%%%%%%%%%%%%%%%%%%%%%%%%%%%%%%%%%%%
%

\section{Discussion}
In Fig.~\ref{Fig6},
we summarize our Seebeck data in Nd-LSCO. 
In Fig.~\ref{Fig6}(a),
we show the evolution of the temperature dependence with doping and identify three regions: 
for $p \geq p^{\star}$, $S/T$ varies weakly with temperature;
for $p < p^{\star}$, it exhibits a large increase at low temperature; 
and for $p < 0.19$, it goes through a maximum before falling at low temperature. 
In Fig.~\ref{Fig6}(b), we plot the normal-state value of $S/T$ measured at $T = 5$~K,
as a function of doping, and again define three regions: 
for $p \geq p^{\star}$, $S/T$ is small, positive and weakly doping-dependent; 
for $p < p^{\star}$, $S/T$ is initially large, positive and roughly constant; 
and for $p < 0.19$, $S/T$ drops with decreasing $p$ to eventually become negative.
In the following, we discuss in more detail this evolution of $S/T$ across the phase diagram of Nd-LSCO.

\subsection{Overdoped metal ($p >$~\pstar)}
At $p = 0.24$, a doping just above \pstar,
the resistivity displays a perfect $T$-linear dependence down to $T \rightarrow 0$~\cite{daou2009,collignon2017},
as reproduced in Fig.~\ref{Fig7}(a).
This is a classic signature of quantum criticality (specifically for an antiferromagnetic QCP in 2D)~\cite{lohneysen2007}, 
suggesting that \pstar~is a quantum critical point.
This inference is strongly supported by the logarithmic dependence of the electronic specific heat in Nd-LSCO at \pstar,
whereby $C_{\rm el}/T \propto$~log$(1/T)$~\cite{michon2019}, 
a second classic signature of the same type of quantum criticality~\cite{lohneysen2007}.
Note that the slope of the $T$-linear resistivity is consistent with the so-called Planckian limit for the scattering rate,
a universal property of cuprates~\cite{legros2019} and other materials near a quantum critical point~\cite{bruin2013}.

At $p = 0.24$, the Seebeck coefficient also displays a logarithmic dependence, whereby $S/T \propto$~log$(1/T)$,
as reported previously for Nd-LSCO~\cite{daou_thermo2009} and Eu-LSCO~\cite{laliberte2011}.
Recently, a log$(1/T)$ dependence was also observed in Bi2201 near \pstar~\cite{lizaire2020}.
This is also considered to be a signature of the same type of quantum criticality~\cite{paul2001}.

\textcolor{black}{In Nd-LSCO, the magnetism detected by neutrons decreases in intensity with doping and becomes quite weak above $p = 0.24$, but some short-range slowly fluctuating magnetism does seem to persist up to roughly $p = 0.26$~\cite{ma2020}.
On the other hand, a slower probe like muons does not see any static magnetism in Nd-LSCO at $p = 0.20$, or above~\cite{nachumi1998}.
As a result, the clear identification of a magnetic quantum critical point by spectroscopy depends on the time scale of the probe.
In that respect, transport and thermodynamic measurements detect the doping at which the magnetism on a timescale that affects the electronic properties.
Note that nuclear magnetic resonance and ultrasound measurements in LSCO, both slower probes than neutrons, find that magnetism ends at \pstar~in that material~\cite{frachet2020}.}

%%%%%%%%%%%%%%% Begin Figure 8 %%%%%%%%%%%%%%%%%%%%%%%%%%%%%%%%%%%%%%%%%%%%%%%%%%%%%%%%%%%%
%
\begin{figure}[t!]
\includegraphics[width=0.47\textwidth]{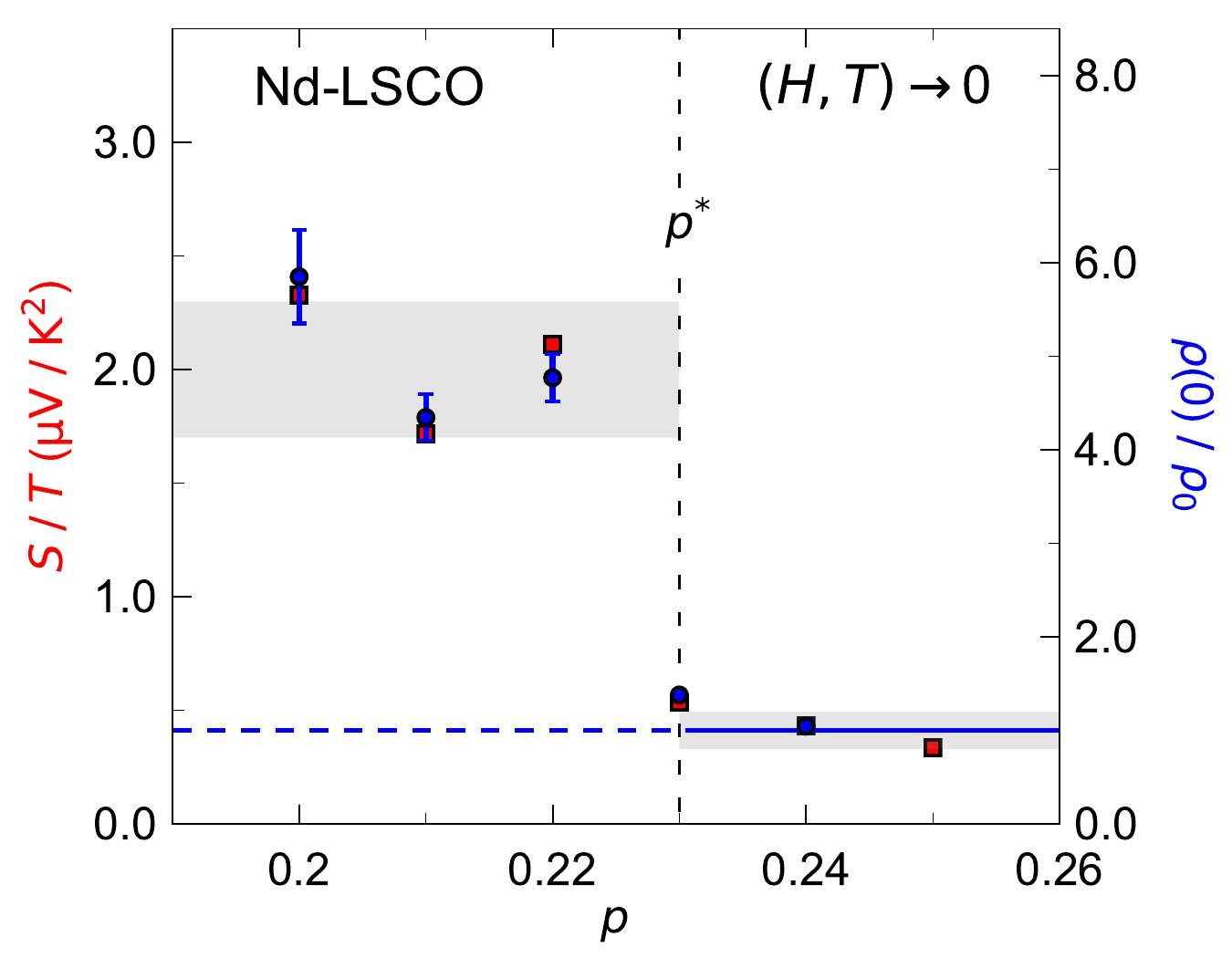}
\caption{
Seebeck coefficient $S/T$ (red squares) and resistivity ratio $\rho(0)/\rho_0$ (blue dots, see main text) as a function of doping in Nd-LSCO. Both quantities are taken in the $H \rightarrow 0$ and $T \rightarrow 0$ limits, and both exhibit a sudden jump at the pseudogap critical point $p^{\star}$, a signature of the underlying change in carrier density.
}
\label{Fig8}
\end{figure}
%
%%%%%%%%%%%%%%% End Figure 8 %%%%%%%%%%%%%%%%%%%%%%%%%%%%%%%%%%%%%%%%%%%%%%%%%%%%%%%%%%%%

\subsection{Pseudogap phase ($p <$~\pstar)}
In Nd-LSCO, the entry into the pseudodap phase at \pstar, 
in the normal state at $T \to 0$,
has been examined via a number of transport measurements. 
The Hall coefficient \RH~and the electrical resistivity $\rho$, which in a simple model would both be
inversely proportional to the carrier density $n$, show a large increase when the doping is reduced below \pstar~\cite{collignon2017}, 
by a factor of about 5. 
The thermal conductivity $\kappa$, which would be proportional to $n$, shows a drop by the same factor upon crossing \pstar, in $\kappa/T$ in the $T=0$ limit (ref.~\onlinecite{michon2018}).
(In other words, the Wiedemann-Franz law is satisfied.)
The present paper now adds the Seebeck coefficient, inversely proportional to $n$~(Eq.~3), which shows a large enhancement at \pstar, by a factor of $\sim 5$ between $p = 0.24$-0.25 and $p = 0.19$-0.22 (Figs.~\ref{Fig4} and \ref{Fig6}(b)).
\textcolor{black}{We stress that Eq.~3 is used here to interpret the large relative change in $S/T$ at \pstar, not to understand the full amplitude and sign of $S/T$, to which many factors may contribute.}

In Fig.~\ref{Fig7}, we compare $S/T$ and the resistivity $\rho$ in Nd-LSCO at two dopings, above and below \pstar.
At $p = 0.22$, $\rho(T)$ displays a pronounced upturn at low temperature~\cite{collignon2017} whose onset, 
taken as the initial deviation of $\rho(T)$ from the high-temperature $T$-linear regime (Fig.~\ref{Fig7}(a)), 
is a standard definition of the pseudogap temperature \Tstar. 
This signature matches the actual pseudogap opening observed in ARPES measurements~\cite{matt2015} (Fig.~\ref{Fig1}), 
and is absent at $p = 0.24$ (Fig.~\ref{Fig7}(a)), where there is no pseudogap~\cite{matt2015}. 
As seen in Fig.~\ref{Fig7}(b), an upturn in $S/T$ is observed at low temperature at $p = 0.22$, relative to the slow growth seen at $p = 0.24$, and it starts roughly at \Tstar.
The two upturns, in $\rho$ and in $S/T$, are two signatures of the pseudogap phase that develops below \Tstar.

%%%%%%%%%%%%%%% Begin Figure 9 %%%%%%%%%%%%%%%%%%%%%%%%%%%%%%%%%%%%%%%%%%%%%%%%%%%%%%%%%%%%
%
\begin{figure}[t!]
\includegraphics[width=0.47\textwidth]{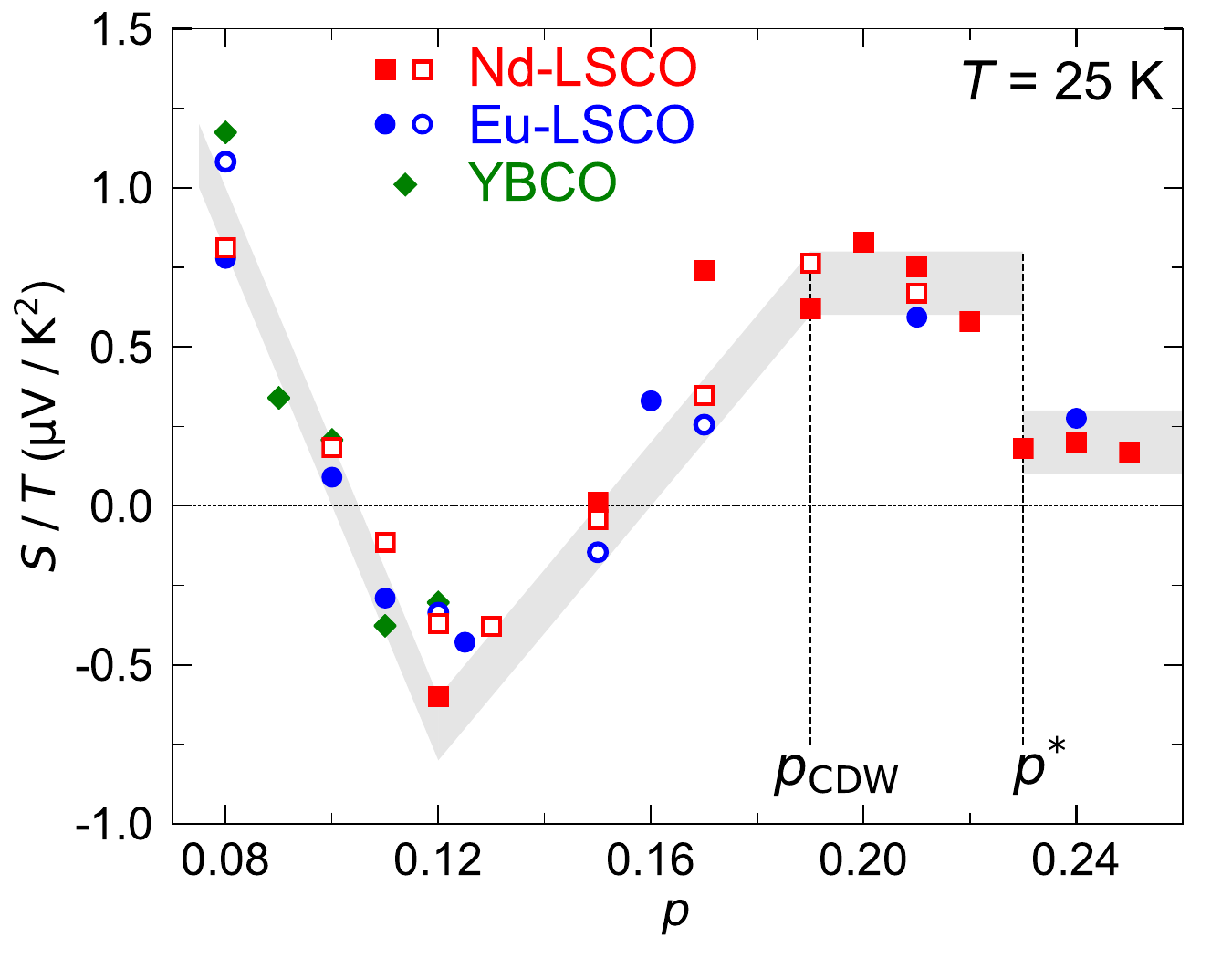}
\caption{
Magnitude of $S/T$ in the normal state at $T = 25$~K for three different cuprates as a function of doping:
Nd-LSCO ($y = 0.4$, solid red squares, this work; $y = 0.6$, open red squares~\cite{hucker1998});
Eu-LSCO ($y = 0.15$, open blue circles~\cite{hucker1998}; $y = 0.2$, solid blue circles~\cite{laliberte2011});
YBCO (solid green diamonds~\cite{laliberte2011}).
The grey band is a guide to the eye.
In addition to the sudden jump in $S/T$ upon entering the pseudogap phase below \pstar,
this shows the universal drop in $S/T$ towards negative values upon entering the CDW phase 
at $p_{\rm CDW}$ 
($\simeq 0.19$ in Nd-LSCO and Eu-LSCO, $\simeq 0.16$ in YBCO).
}
\label{Fig9}
\end{figure}
%
%%%%%%%%%%%%%%% End Figure 9 %%%%%%%%%%%%%%%%%%%%%%%%%%%%%%%%%%%%%%%%%%%%%%%%%%%%%%%%%%%%

In the $T = 0$ limit, the normal-state values of $\rho$ and $S/T$ display a comparable 5-fold increase between $p = 0.24$ and $p = 0.22$. 
Since (in a simple model) $S/T \propto m^{\star}/n$, $\rho \propto m^{\star}/(n\tau)$, and \RH~$\propto 1 / n$,
where $1/\tau$ is the scattering rate, this suggests that it is a change in carrier density that causes the jump of comparable magnitude 
in the three quantities across \pstar, 
with negligible change in $1/\tau$. 
Magnetoresistance measurements provide independent evidence that $1/\tau$ does not change significantly across \pstar,
since the amplitude of the orbital magnetoresistance (controlled by $\omega_c \tau$) is very similar on both sides of \pstar,
whether for $\rho_a$~\cite{collignon2017} or for $\rho_c$~\cite{fang2020}, 
despite the large change in the absolute value of $\rho_a(T = 0)$~\cite{collignon2017} and $\rho_c(T = 0)$~\cite{cyr-choiniere2010}.

For a more quantitative comparison between the two coefficients, 
the effect of the magnetic field should be accounted for in both $S/T$ and $\rho$. 
An estimate of the zero-field resistivity below \Tc, from field sweeps and temperature sweeps, was detailed in ref.~\onlinecite{collignon2017} 
and here we use those values. 
In Fig.~\ref{Fig8},
we plot the values of $S/T$ in the $H \rightarrow 0$ and $T \rightarrow 0$ limits as a function of doping across \pstar. 
For the resistivity, we plot the ratio $\rho(0) / \rho_0$ as a function of doping, 
where $\rho(0)$ is the value of $\rho$ in the $H \rightarrow 0$ and $T \rightarrow 0$ limits 
and $\rho_0$ is the value that the resistivity would have in the absence of the pseudogap. 
As revealed by a study where the pseudogap is turned on and off by pressure in Nd-LSCO~\cite{doiron-leyraud2017}, 
$\rho_0$ can be extracted by extending the high-temperature $T$-linear regime to $T = 0$, 
as illustrated for $p = 0.22$ by the dashed-dotted line in Fig.~\ref{Fig7}(a). 
The rationale for taking this ratio is that it removes the effect of disorder, which is different for different samples.
For example, we can see that $\rho_0$ is slightly larger in our sample with $p = 0.22$ than it is in our sample with $p = 0.24$.
(By contrast, the Seebeck coefficient does not depend on the disorder level of a particular sample.)
As seen in Fig.~\ref{Fig8}, 
both $S/T$ and $\rho(0)/\rho_0$ display the same sudden jump at \pstar~and track each other below \pstar. 
Given that in a simple model both $\rho$ and $S/T$ vary as $1 / n$,
the parallel jump in $\rho$ and $S/T$ below \pstar~is further evidence for a drop in carrier density, 
initially inferred from the Hall coefficient~\cite{badoux2016,collignon2017}.
A large positive $S/T$ at low $T$ is also observed in other cuprates below \pstar,
such as Bi2201 and Bi2212 \cite{kondo2005}.

Upon entering the pseudogap phase, the electronic specific heat $C_{\rm el}$ of cuprates drops.
In Nd-LSCO, the normal-state $C_{\rm el}/T$ at $T = 2$~K decreases by a factor $\sim 2$ when going from $p = 0.24$
to $p = 0.20$ \cite{michon2019}.
So according to Eq.~3, this only reinforces our inference that the carrier density must drop.
Of course, Eq.~3 is an oversimplification and a proper theory of the Seebeck coefficient
in cuprates is needed to understand in detail its behavior, whether above or below \pstar.
To highlight this need, let us mention that a calculation of $S$ based on the band structure of Nd-LSCO yields a negative value above \pstar~\cite{verret2017}, in contrast with the measured positive value.

\subsection{Charge-density-wave phase ($p <$~\pcdw)}
Early Seebeck measurements on Nd-LSCO and Eu-LSCO revealed that $S$ is negative at low $T$
for dopings close to $p = 0.12$~\cite{nakamura1992,hucker1998} (see open symbols in Fig.~\ref{Fig9}).
Later studies showed that a negative $S$ at low $T$ is also observed at $p \simeq 0.12$ in
LBCO~\cite{li2007},
YBCO~\cite{chang2010},
Hg1201~\cite{doiron-leyraud2013},
and LSCO~\cite{badoux2016a}
showing that it is a universal property of hole-doped cuprates.
The negative $S$ at low $T$ came to be understood as a consequence of a Fermi-surface reconstruction
caused by the onset of CDW order~\cite{taillefer2009,laliberte2011},
whose other signatures are a negative Hall coefficient~\cite{leboeuf2007,doiron-leyraud2013} 
and low-frequency quantum oscillations~\cite{doiron-leyraud2007,barisic2013}.
In Fig.~\ref{Fig9},
we gather together Seebeck data on Nd-LSCO, Eu-LSCO and YBCO by plotting the value of $S/T$ at $T = 25$~K vs doping.
The agreement between early data on Nd-LSCO and our own data is excellent,
as is the agreement between early and later data on Eu-LSCO.
The fact that $S/T$ has its minimum (most negative) value at $p = 0.12$ is consistent with the fact that CDW order
is strongest at $p = 0.12$ in Nd-LSCO, Eu-LSCO and YBCO (and LBCO and LSCO).

In Fig.~\ref{Fig6},
we define \Tmax~to be the temperature below which $S/T$ starts to drop towards negative values.
The values of \Tmax~are plotted vs $p$ in the phase diagram of Fig.~\ref{Fig1},
where we also show the corresponding values for the closely related material Eu-LSCO (taken from~\cite{laliberte2011}).
The fact that \Tmax~matches \Tcdw, the onset temperature for CDW order detected by x-ray diffraction
in Nd-LSCO\textcolor{black}{~\cite{zimmermann1998,niemoller1999,gupta2020}} and Eu-LSCO~\cite{fink2011},
confirms that the drop in $S/T$ is caused by CDW order (Fig.~\ref{Fig1}).
We see that \Tmax~decreases with doping, and vanishes close to $p = 0.19$.
We therefore find no evidence of the usual effect of CDW order on the transport properties of Nd-LSCO and Eu-LSCO beyond \pcdw~$= 0.19$ (Figs.~\ref{Fig1},~\ref{Fig6}(b) and~\ref{Fig9}).
\textcolor{black}{This is consistent with a recent x-ray study of Nd-LSCO that finds no CDW order at $p = 0.18$ and 0.19, down to 20~K~\cite{gupta2020}.}

Note that the start of the CDW phase on the low doping side can also be detected using Seebeck data.
In Eu-LSCO and YBCO, there is no trace of a drop in $S/T$ as $T \to 0$ for $p = 0.08$~\cite{laliberte2011}.
The same is true for Nd-LSCO~\cite{hucker1998}.
We infer that the critical doping for the onset of the CDW phase in these three materials is $p = 0.08$ (Fig.~\ref{Fig9}).
This is nicely consistent with x-ray diffraction studies that show no CDW modulations at $p < 0.08$
in these same cuprates~\cite{fink2011,blanco-canosa2014}.

\section{Summary}

We have used measurements of the Seebeck coefficient $S$
to investigate the phase diagram of the cuprate superconductor Nd-LSCO, 
in the normal state as $T \to 0$,
reached by applying a magnetic field large enough to suppress superconductivity.
We identify three regions (Figs.~\ref{Fig6}(b) and \ref{Fig9}):
1) $p >$~\pstar, outside the pseudogap phase, where $S$ is positive and $S/T$ is small;
2) \pcdw~$< p <$~\pstar, inside the pseudogap phase but without CDW order, where $S$ is positive and $S/T$ is large;
3) $p <$~\pcdw, inside the CDW phase, where $S/T$ decreases as $T \to 0$.
The large and sudden increase in $S/T$ that occurs upon crossing below the pseudogap critical doping \pstar~is further
evidence for a drop in carrier density, as also inferred from the drop in Hall number and in conductivity.
The sudden change from increasing to decreasing $S/T$ as $T \to 0$ in going from $p = 0.19$ to $p = 0.17$
allows us to identify \pcdw~$= 0.19$ as the upper bound for the end of the CDW phase as seen by transport.
This reveals that \pcdw~is well separated from \pstar, as in the case of YBCO and LSCO.

\section{Acknowledgements}

We thank S.~Fortier for his assistance with the experiments.
L.T. and B.D.G. acknowledge support from the Canadian Institute for Advanced Research 
(CIFAR) as CIFAR Fellows
and funding from the Natural Sciences and Engineering Research Council of Canada.
L.T. acknowledges funding from the Institut Quantique, 
the Fonds de Recherche du Qu\'ebec -- Nature et Technologies (FRQNT), 
the Canada Foundation for Innovation (CFI), 
and a Canada Research Chair.
This research was undertaken thanks in part to funding from the Canada First Research Excellence Fund
and the Gordon and Betty Moore Foundation's EPiQS Initiative (Grant GBMF5306 to L.T.).
J.S.Z. was supported by NSF MRSEC under Cooperative Agreement No. DMR-1720595.
This work was supported by HFML-RU/NWO, a member of the European Magnetic Field Laboratory (EMFL).

$^{\star \star}$ These authors contributed equally to this work.

%%%%%%%%%%%%%%%%%   REFERENCES

%\bibliography{Collignon_references}


\onecolumngrid

\pagebreak

\setcounter{figure}{0}
\renewcommand{\thefigure}{S\arabic{figure}}

\begin{center}
{\Large Supplementary Material for\\
``Thermopower across the phase diagram of the cuprate La$_{1.6-x}$Nd$_{0.4}$Sr$_x$CuO$_4$ :\\
signatures of the pseudogap and charge-density-wave phases"}
\end{center}

%
%%%%%%%%%%%%%%% Begin Figure 1 %%%%%%%%%%%%%%%%%%%%%%%%%%%%%%%%%%%%%%%%%%%%%%%%%%%%%%%%%%%%
%
\begin{figure}[h]
\includegraphics[scale=0.8]{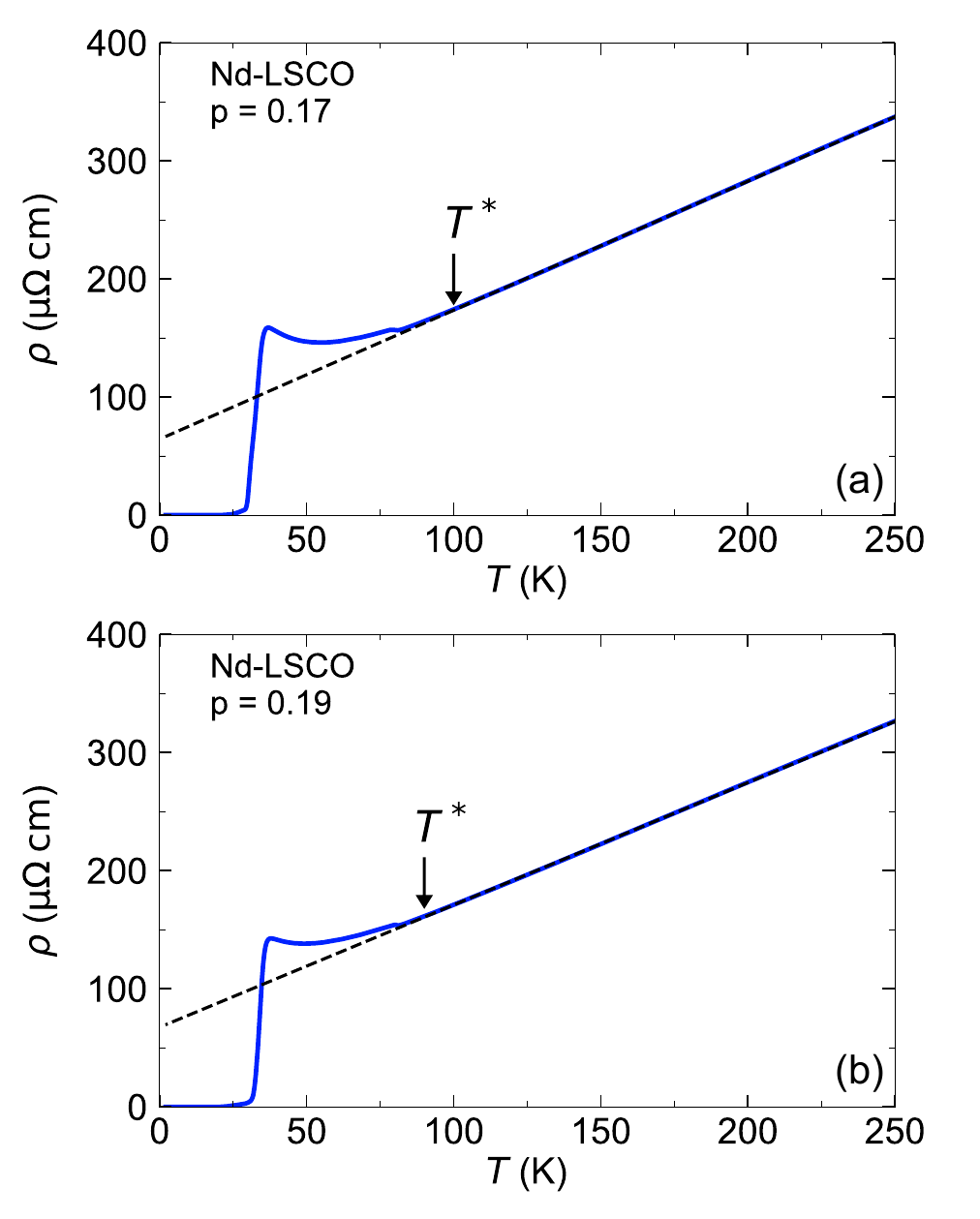}
\caption{
Electrical resistivity $\rho$ as a function of temperature for Nd-LSCO at $p = 0.17$ (a) and $p = 0.19$ (b), showing the linear-$T$ behavior at high temperature (black dashed lines) and the departure from that regime at the pseudogap temperature $T^{\star}$. The values of $T^{\star}$ are 100~$\pm 10$~K at $p = 0.17$ and 90~$\pm 10$~K at $p = 0.19$.
}
\label{FigS1}
\end{figure}
%
%%%%%%%%%%%%%%% End Figure 1 %%%%%%%%%%%%%%%%%%%%%%%%%%%%%%%%%%%%%%%%%%%%%%%%%%%%%%%%%%%%
%
%
%%%%%%%%%%%%%%% Begin Figure 2 %%%%%%%%%%%%%%%%%%%%%%%%%%%%%%%%%%%%%%%%%%%%%%%%%%%%%%%%%%%%

\begin{figure}[t]
\includegraphics[scale=0.7]{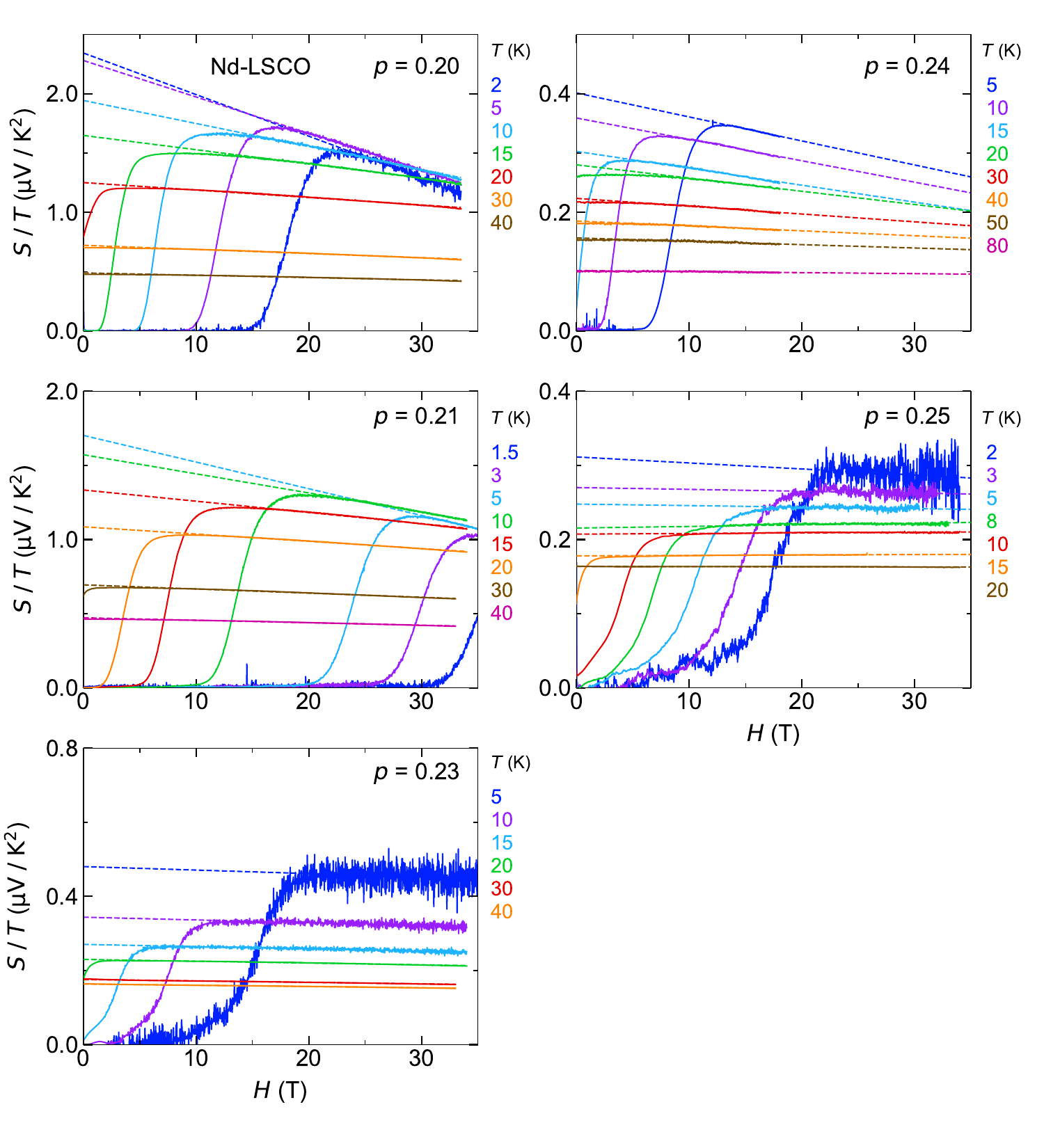}
\caption{
Seebeck coefficient $S$ plotted as $S/T$ as a function of magnetic field for Nd-LSCO at dopings as labeled, at temperatures as indicated. The dashed lines are linear fits to the normal state $S/T (H)$, whose extrapolation to $H = 0$ provide values of $S/T$ in the absence of a magneto-Seebeck effect (see main text).
}
\label{FigS2}
\end{figure}
%
%%%%%%%%%%%%%%% End Figure 2 %%%%%%%%%%%%%%%%%%%%%%%%%%%%%%%%%%%%%%%%%%%%%%%%%%%%%%%%%%%%
%
%
%%%%%%%%%%%%%%% Begin Figure 2 %%%%%%%%%%%%%%%%%%%%%%%%%%%%%%%%%%%%%%%%%%%%%%%%%%%%%%%%%%%%

\begin{figure}[t]
\includegraphics[scale=0.6]{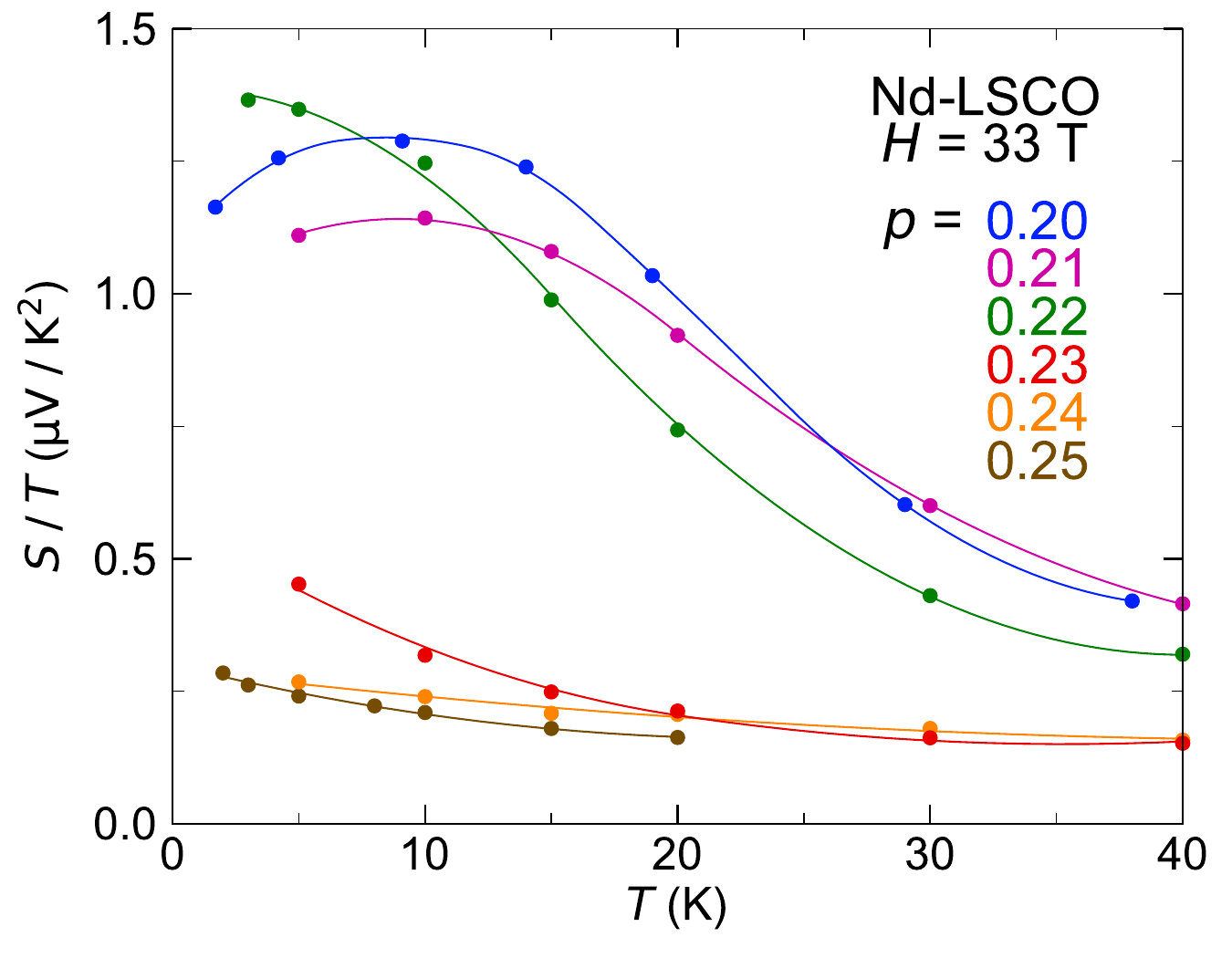}
\caption{
Seebeck coefficient $S/T$ as a function of temperature in Nd-LSCO at dopings as indicated, in $H = 33$~T.
}
\label{FigS3}
\end{figure}
%
%%%%%%%%%%%%%%% End Figure 2 %%%%%%%%%%%%%%%%%%%%%%%%%%%%%%%%%%%%%%%%%%%%%%%%%%%%%%%%%%%%

\end{document}